%% file: main.tex
\documentclass[runningheads]{llncs}
\usepackage[T1]{fontenc}
\usepackage{graphicx}
\usepackage{cite}
\usepackage{amsmath,amssymb,amsfonts, mathtools}
\usepackage{algorithmic}
\usepackage{textcomp}
\usepackage{soul}
\usepackage{xcolor}
\usepackage{hyperref}

\newcommand{\sname}{\texttt{InfoHier}}

\begin{document}
\title{\sname: Hierarchical Information Extraction via Encoding and Embedding}
%
%
\author{Tianru Zhang\inst{1} \and
Li Ju\inst{1} \and
Prashant Singh\inst{1,2} \and
Salman Toor\inst{1}}
%
\authorrunning{T. Zhang et al.}
%
\institute{Uppsala University, Uppsala, Sweden\\
\email{tianru.zhang@it.uu.se}
\and
Science for Life Laboratory, Sweden\\
}
\maketitle              
\vspace{-0.2in}
\begin{abstract}
Analyzing large-scale datasets, especially involving complex and high-dimensional data like images, is particularly challenging. While self-supervised learning (SSL) has proven effective for learning representations from unlabeled data, it typically focuses on flat, non-hierarchical structures, missing the multi-level relationships present in many real-world datasets. Hierarchical clustering (HC) can uncover these relationships by organizing data into a tree-like structure, but it often relies on rigid similarity metrics that struggle to capture the complexity of diverse data types. To address these we envision \sname{}, a framework that combines SSL with HC to jointly learn robust latent representations and hierarchical structures. This approach leverages SSL to provide adaptive representations, enhancing HC's ability to capture complex patterns. Simultaneously, it integrates HC loss to refine SSL training, resulting in representations that are more attuned to the underlying information hierarchy. \sname{} has the potential to improve the expressiveness and performance of both clustering and representation learning, offering significant benefits for data analysis, management, and information retrieval.

\end{abstract}
%
%
\section{Introduction}
\label{sec:intro}
\input{sections/1-intro}

\section{Background}
\label{sec:bg}
\input{sections/2-bg}

\section{Method}
\label{sec:method}
\input{sections/3-method}

\section{Preliminary Results and Applications}
\label{sec:result}
\input{sections/4-results}
\section{Conclusion and Future Directions}
\label{sec:prospects}
\input{sections/5-conclusion}


%
%
%
\bibliographystyle{splncs04}
\bibliography{ref}

\end{document}

%% file: sections/1-intro.tex
In the present era of big data, vast amounts of information are generated every second, posing significant challenges for data analysis \cite{big-data, marx2013big, big-data-analy}. 
Manual analysis is no longer feasible due to the large scale and high velocity of data generation in domains like bio-science, finance, and autonomous systems. This demand has fueled the rise of machine learning, particularly deep learning, which excels at analyzing complex data.

While supervised learning achieves state-of-the-art performance, it relies heavily on large, labeled datasets, which are typically costly to obtain. This is especially problematic in fields like biomedical imaging, where labeling requires expert knowledge. Self-supervised learning (SSL) has emerged as an alternative, learning from unlabeled data by solving pretext tasks to generate useful representations for downstream tasks.

\begin{figure}
    \centering
    \includegraphics[width=0.9\linewidth]{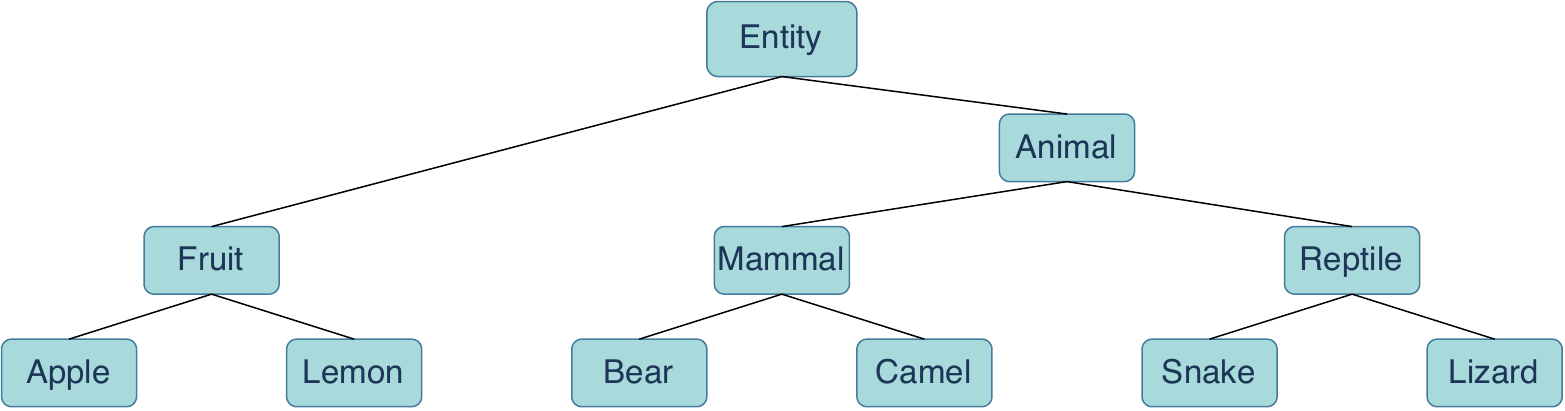}
    \caption{Illustration of the inherent hierarchical structure of categories in ImageNet~\cite{imagenet}.}
    \label{fig:imagenet}
\end{figure}

Despite their success, SSL methods typically neglect the hierarchical structure present in many real-world datasets. For instance, ImageNet~\cite{imagenet} contains a hierarchy of categories (e.g., animals and fruits in \autoref{fig:imagenet}), yet most SSL approaches focus on flat representations, ignoring this multi-level structure.

\begin{figure}
    \centering
    \includegraphics[width=0.9\linewidth]{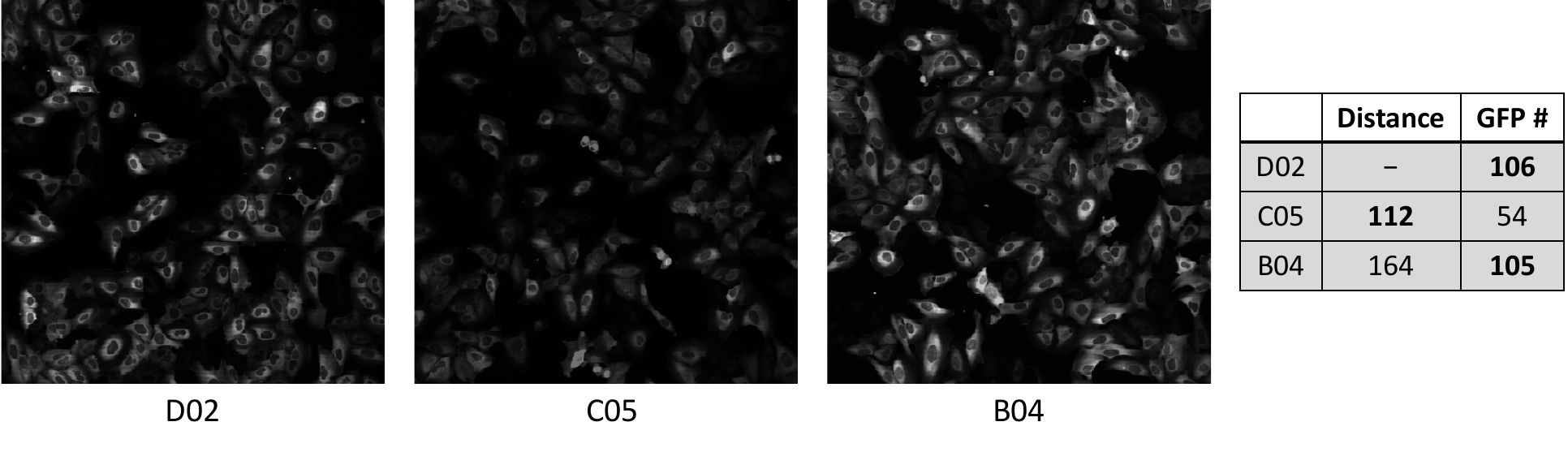}
    \caption{Images with very similar GFP(Green Fluorescent Protein)-in-cytoplasm counts are not identified by the distance metric.}
    \label{fig:dist_fail}
\end{figure}

On the other hand, hierarchical clustering (HC) is a well-established unsupervised learning technique that can uncover such relationships by organizing data into tree-like structures. However, HC relies on rigid similarity-based metrics (linkages), which often struggle with high-dimensional, complex data, as well as be sensitive to perturbations. For example, as illustrated in \autoref{fig:dist_fail}, the distance metric fails to correctly identify images from the BBBC013 dataset \cite{BBBC013} that should be grouped as similar. Image D02 and B04 exhibit highly similar feature (GFP-in-cytoplasm counts), but due to the limitations of the distance metric, they are not clustered together.

To address these limitations, we propose \sname{}, an innovative framework that combines the strengths of SSL with hierarchical clustering. By jointly training SSL and HC, \sname{} learns representations that reflect the underlying hierarchy of data, enhancing both clustering and representation learning for complex datasets.

%% file: sections/2-bg.tex

This section provides an overview of foundational techniques that guide our proposed framework, focusing on SSL for effective representation learning from unlabeled data and HC for uncovering nested structures in complex datasets.

\subsection{Self-Supervised Learning}

Self-supervised learning is designed to extract informative representations from vast unlabeled datasets by using pretext tasks, that encourage the model to learn meaningful data features without explicit labels. SSL methods generally fall into three categories: generative, contrastive, and adversarial~\cite{SSL-TKDE, cookbookssl}.
Generative SSL approaches, such as autoencoders, use an encoder-decoder framework to reconstruct input data by learning relevant latent features. They are trained via a reconstruction loss that measures the input-output similarity. Extensions to autoencoders like denoising autoencoders~\cite{denoisingAE} and masked autoencoders~\cite{maskedAE} enhance robustness by introducing partial corruption to input data, requiring the model to reconstruct the missing information. While generative methods are effective for data generation and imputation tasks, they often lack the discriminative power needed for other tasks.

Contrastive SSL methods, on the other hand, emphasize relational learning by maximizing similarity between positive pairs (e.g., augmented views of the same sample) and minimizing it for negative pairs (distinct samples). Deep metric learning methods like CPC~\cite{cpc} and SimCLR~\cite{simclr} leverage augmentations and similarity objectives to produce discriminative latent representations. Further contrastive approaches include self-distillation methods such as BYOL~\cite{byol}, SimSiam~\cite{simsiam}, and DINO~\cite{dino}, which train models by mapping different views of an input through paired encoders, and use a predictor to align one view with the other. Canonical Correlation Analysis (CCA)-inspired methods like VICReg~\cite{vicreg}, Barlow Twins~\cite{barlow}, and W-MSE~\cite{w-mse} enhance similarity by aligning cross-covariance matrices, promoting feature diversity and generalization. To unify these diverse objectives, researchers have formulated SSL losses as differentiable functions within a single framework~\cite{unifiedCL}, and some studies have also provided theoretical foundations from spectral graph perspectives~\cite{provablessl, spectralCL}. 

In parallel, clustering-based SSL methods like DeepCluster~\cite{DeepCluster}, JULE~\cite{JULE}, SeLa~\cite{SeLa}, and SwAV~\cite{swav} combine clustering with representation learning by assigning pseudo-labels to data clusters, guiding the network to discover latent data structures without labeled supervision. However, these methods typically use flat clustering (e.g., k-means and nearest neighborhood), which overlooks potential hierarchical structures inherent in complex datasets.

\subsection{Hierarchical Clustering}

Hierarchical clustering provides a method for organizing data into nested clusters, revealing multi-level relationships across different groups. Unlike flat clustering methods, HC builds tree-like structures that capture hierarchies and relationships across various levels. Traditional HC approaches rely on predefined distance metrics and linkage-based algorithms, which struggle with complex data variations and are sensitive to factors like noise or transformations.

Recent work has focused on improving HC efficacy, such as through the use of the Dasgupta loss~\cite{dasguptaHC}, an optimization-based metric that rewards close intra-cluster points while penalizing distant ones. However, directly minimizing the Dasgupta loss remains challenging, especially for large datasets. To address this, gradient-based hierarchical clustering approaches have emerged, enabling optimization via gradient descent in the space of hierarchical structures~\cite{UFit, gHHC, HypHC}. By using hyperbolic embeddings and continuous relaxations, these methods facilitate HC’s integration with neural networks, allowing end-to-end learning of both data representations and hierarchical structures.

\subsection{Research Focus}

Our research targets current limitations in extracting informative representations and underlying structures from complex, unlabeled datasets. In this context, integrating SSL with HC is promising. SSL, especially contrastive methods, generates robust latent representations that refine similarity measurements in HC. Our proposed \sname{} framework jointly trains SSL and HC, allowing HC to leverage refined latent features while SSL also gains from hierarchical structure learning. This integration shows potential for hierarchical representation learning within complex datasets with intricate structure.

%% file: sections/3-method.tex

\begin{figure}[!h]
    \centering
    \includegraphics[width=\columnwidth]{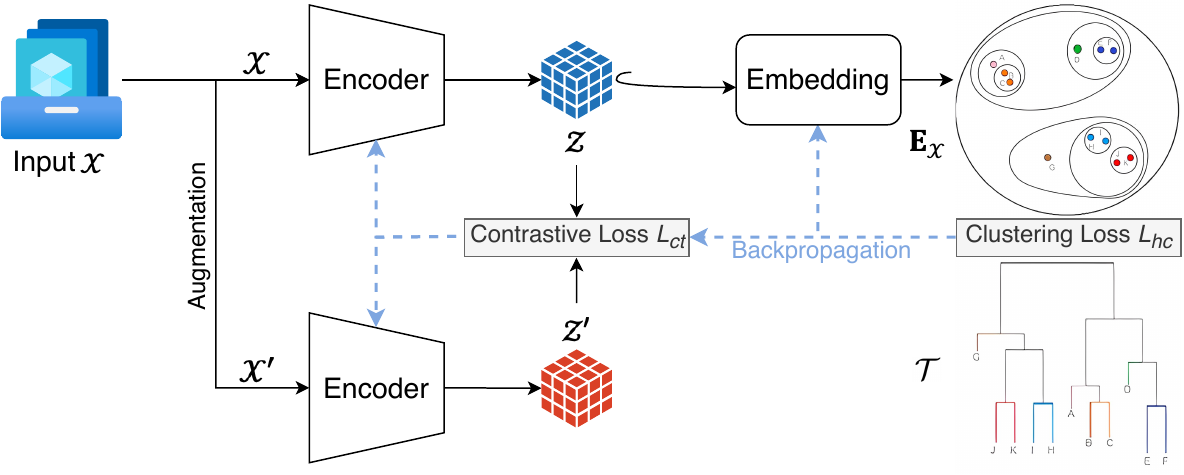}
    \caption{Structural overview of \sname, where solid lines are the data flow and dashed lines represent the gradient flow.}
    \label{fig:structure}
\end{figure}

Our end-to-end objective can be described as follow: For a dataset $\mathcal{X} \in\mathbb{R}^{N\times d}$ containing $N$ data points with dimension $d$, extract the hierarchical structure in the form of a rooted binary tree $\mathcal{T}$ with $N$ leaves, such that each leaf corresponds to a datapoint in $\mathcal{X}$, and intermediate nodes represent clusters.

To address the limitation of SSL and HC as discussed in earlier sections, we propose a joint framework as shown in \autoref{fig:structure}. The raw input $\mathcal{X}$ is first encoded into a latent representation $\mathcal{Z}$ through a pre-trained encoder network, then embedded into a two-dimensional hyperbolic space $\mathbb{B}_2$ using an embedding map $\mathbf{E}_\mathcal{X}$, and eventually decoded to a tree $\mathcal{T}$ representing the hierarchical structure of the input. To improve the hierarchical clusters, we employ the continuous version of the Dasgupta loss as the HC loss to train the embedding network. Moreover, the contrastive loss is combined to further fine-tune the encoder network, in order to train the encoder towards more robust latent representations that are aware of the underlying hierarchy. In the following paragraphs, we show the detailed formulas of each loss and how they are combined to serve the joint training of the framework.

We start with the gradient-based HC training, reformulating the Dasgupta HC loss into a differentiable form. The Dasgupta cost function is motivated by encouraging similar points to be nearby in the tree structure, formed as $\mathcal{L}_{\text{Das}}=\sum_{(u,v)\in\mathcal{T}}w(u,v)\cdot \#_l(\texttt{lca}(u,v)) $, where $u,v$ are two leaf nodes in tree $\mathcal{T}$ representing two data points in $\mathcal{X}$, $w(\cdot,\cdot)$ is the similarity between two nodes, $\texttt{lca}(\cdot, \cdot)$ denotes the lowest common ancestor (lca) of two nodes, and $\#_l(\cdot)$ denotes the number of leaves in the sub-tree of lca. By continualizing $\#_l(\texttt{lca}(u,v))$ using the hyperbolic distance and reformulating the Dasgupta loss into a triplet form, the HC loss is expressed as follows:
\begin{equation}
\label{hc_loss}
    \mathcal{L}_{\text{hc}}(\mathbf{E}_\mathcal{X}, w)=\sum_{i,j,k}(w_{ij}+w_{jk}+w_{ik}-w_{ijk}(\mathbf{E}_\mathcal{X};w)),
\end{equation}
where $\mathbf{E}_\mathcal{X}\in\mathbb{B}_2$ denotes the embedding of $\mathcal{X}$, $w_{ij}$ is the similarity between data point $i$ and $j$, and $w_{ijk}(\mathbf{E}_\mathcal{X};w)$ is the triplet similarity that is determined by the tree structure, or equivalently the embedding $\mathbf{E}_\mathcal{X}$. With this continuous adaption, the embedding network can be trained by minimizing the HC loss with gradient-based methods, enabling learning better embeddings $\mathbf{E}_\mathcal{X}$ and thus resulting in a more accurate tree.

Thereafter, we introduce the SSL component, which aims to train the encoder network for hierarchy-aware latent representations. Despite the various SSL methods, including but not limited to approaches discussed in \autoref{sec:bg}, they share a common principle: to define a discriminative loss to teach models to distinguish individual input from each other.
Various contrastive losses can be unified in the following concise form as \cite{unifiedCL} presents:
\begin{equation}
\label{ct_loss}
    \mathcal{L}_\text{ct}(\theta)=\sum_{i=1}^N \phi(\sum_{j\neq i}\psi(\Vert z_i-z'_i\Vert_2^2 - \Vert z_i-z_j\Vert_2^2)),
\end{equation}
where $\phi$ and $\psi$ are monotonously increasing and differentiable scalar functions, $\theta$ is the trainable parameters of model, $N$ is the number of data points in input $\mathcal{X}$, and $z_i$ are representations with $z'_i$ as the positive pairs from augmented inputs and $z_j$ as the negative pairs. 

Finally, we combine the two losses into a joint objective:
\begin{equation}
\label{joint_loss}
    \mathcal{L}=\lambda_{\text{ct}}\cdot\mathcal{L}_{\text{ct}}+\lambda_{\text{hc}}\cdot\mathcal{L}_{\text{hc}},
\end{equation}
where $\lambda_{\text{ct}}$ and $\lambda_{\text{hc}}$ are loss balancing hyper-parameters. This integrated loss guides the encoder to produce informative representations that align with the hierarchical structure, resulting in a more robust and discriminative encoder.

%% file: sections/4-results.tex
\begin{figure}[t]
    \centering
    \includegraphics[width=\linewidth]{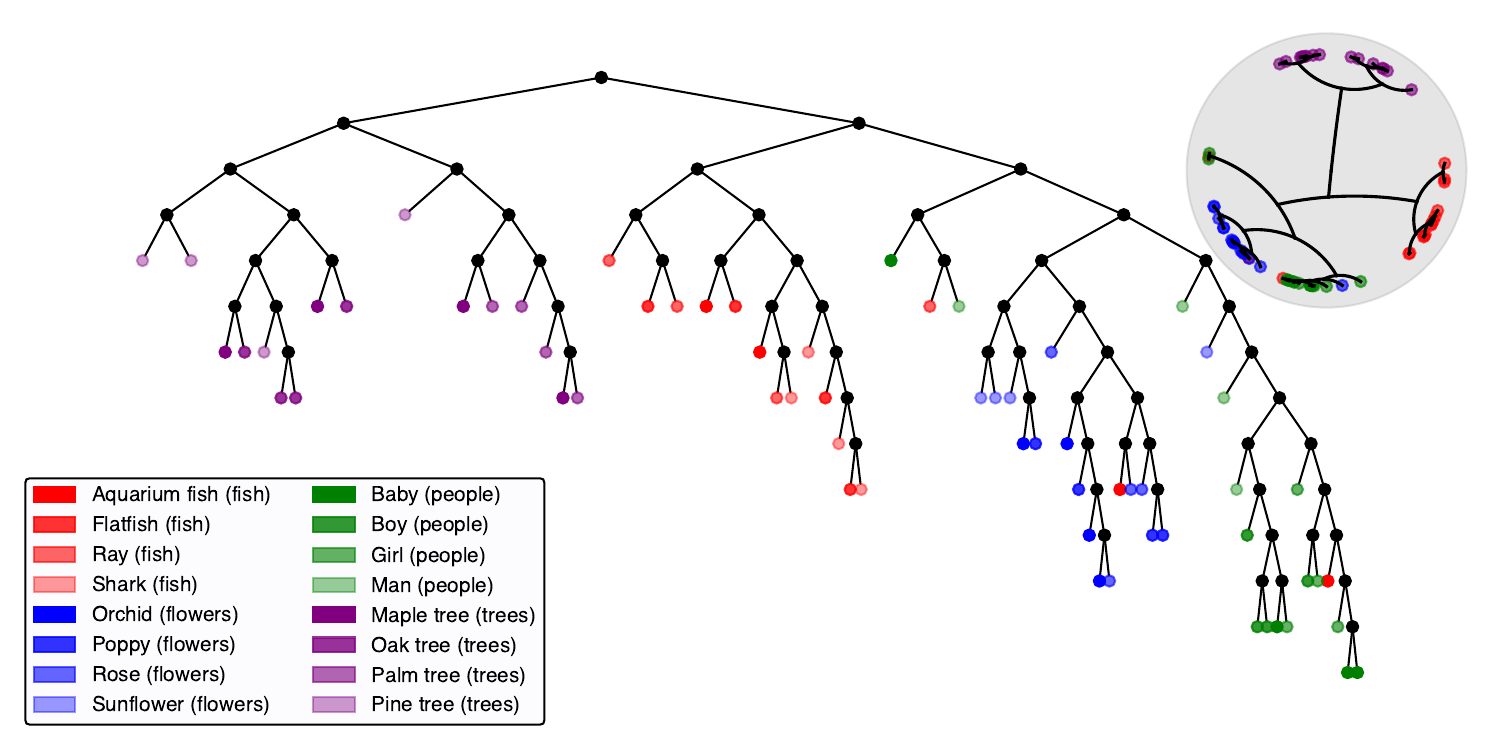}
    \caption{Visualization of the trained framework on 64 samples from the CIFAR100 dataset: Four images are sampled from each of the  16 classes, which can be grouped into four superclasses, denoted by different colors. On top right, the hierarchical structure is visualized on the original hyperbolic space.}
    \label{fig:results}
\end{figure}

In this section, we demonstrate the efficacy of the \sname{} framework with an example implementation. For this experiment, we use SimCLR \cite{simclr} with a ResNet-18 model as the encoder, and a two-dimensional embedding layer as the embedding map. The CIFAR100 \cite{cifar100} dataset is used, and the encoder network is trained using the joint loss function (\ref{joint_loss}) combining the HC loss (\ref{hc_loss}) and NT-Xent loss (a variation of (\ref{ct_loss})). The resulting hierarchical structure is visualized in \autoref{fig:results}. This shows that \sname{} effectively uncovers the hidden structure within the CIFAR100 data, grouping most samples of the same superclass into distinct clusters \emph{without any external label information}. Moreover, beyond human-defined labels, the framework captures intrinsic hierarchical structures in the latent space of the image data, relying solely on the sample representations.

Based on the preliminary result, we envision several practical applications:
In data clustering, \sname{} enables categorization at customizable levels of granularity, which could be beneficial for nuanced categorization.
The hierarchical structure can also support balanced data sampling, providing a way to address imbalanced datasets through structured data groupings. 
\sname{} might also enhance vector databases by facilitating hierarchical indexing \cite{hier_index, vectordb}, improving retrieval efficiency and relevance in product image collections or recommendation systems.
For data storage management, \sname{} can potentially assist in data placement challenges in multi-tiered storage systems \cite{hsmrl, eswa, vldb}, such as identifying high-priority data suitable for faster storage to reduce system latency.

Additionally, \sname{}’s latent structure may support organization in knowledge graph databases \cite{zhang-etal-2018-knowledge, HAKE} and content archives by forming layered, searchable data hierarchies, which could enhance access in large repositories such as scientific archives or multimedia collections. 
It can also streamline the nearest-neighbor search process by narrowing down categories \cite{graph_embd, HNSW}, potentially benefiting recommendation systems and large search infrastructures. 
These applications illustrate how \sname{}’s integration of SSL and HC could address practical challenges in data curation, storage management, and information retrieval, offering a vision of scalable and structured data analysis solutions.

%% file: sections/5-conclusion.tex
In this paper, we introduce \sname{}, a joint framework that integrates SSL and HC to learn informative, hierarchy-aware representations and construct hierarchical structures of inputs without requiring any labels. While the framework is demonstrated on a large-scale problem achieving encouraging results, there is scope for further optimization as future work.
For instance, balancing the different convergence rates of the encoder and embedding networks remains an open question. This may require exploring more sophisticated multi-objective optimization techniques to fine-tune loss-balancing hyperparameters. Scalability is another concern, as incorporating new data points dynamically alters the hierarchical structure. However, with hyperbolic embeddings, \sname{} already offers greater scalability over traditional methods. 

We believe that this approach combining SSL and HC paves the way for more advanced techniques in data analysis and information retrieval, offering richer representations and more structured insights. Ongoing work includes refining the joint loss strategy and conducting comprehensive experiments across diverse use cases. These efforts, along with essential theoretical proofs, are planned for future publication to further substantiate the framework’s effectiveness.

%% file: main.bbl
\begin{thebibliography}{10}
\providecommand{\url}[1]{\texttt{#1}}
\providecommand{\urlprefix}{URL }
\providecommand{\doi}[1]{https://doi.org/#1}

\bibitem{cookbookssl}
Balestriero, R., Ibrahim, M., Sobal, V., Morcos, A., Shekhar, S., Goldstein, T., Bordes, F., Bardes, A., Mialon, G., Tian, Y., Schwarzschild, A., Wilson, A.G., Geiping, J., Garrido, Q., Fernandez, P., Bar, A., Pirsiavash, H., LeCun, Y., Goldblum, M.: A cookbook of self-supervised learning (2023), \url{https://arxiv.org/abs/2304.12210}

\bibitem{vicreg}
Bardes, A., Ponce, J., LeCun, Y.: Vicreg: Variance-invariance-covariance regularization for self-supervised learning. arXiv preprint arXiv:2105.04906  (2021)

\bibitem{DeepCluster}
Caron, M., Bojanowski, P., Joulin, A., Douze, M.: Deep clustering for unsupervised learning of visual features. In: Proceedings of the European conference on computer vision (ECCV). pp. 132--149 (2018)

\bibitem{swav}
Caron, M., Misra, I., Mairal, J., Goyal, P., Bojanowski, P., Joulin, A.: Unsupervised learning of visual features by contrasting cluster assignments. Advances in neural information processing systems  \textbf{33},  9912--9924 (2020)

\bibitem{dino}
Caron, M., Touvron, H., Misra, I., J{\'e}gou, H., Mairal, J., Bojanowski, P., Joulin, A.: Emerging properties in self-supervised vision transformers. In: Proceedings of the IEEE/CVF international conference on computer vision. pp. 9650--9660 (2021)

\bibitem{HypHC}
Chami, I., Gu, A., Chatziafratis, V., R\'{e}, C.: From trees to continuous embeddings and back: hyperbolic hierarchical clustering. In: Proceedings of the 34th International Conference on Neural Information Processing Systems. NIPS '20, Curran Associates Inc., Red Hook, NY, USA (2020)

\bibitem{simclr}
Chen, T., Kornblith, S., Norouzi, M., Hinton, G.: A simple framework for contrastive learning of visual representations. In: International conference on machine learning. pp. 1597--1607. PMLR (2020)

\bibitem{simsiam}
Chen, X., He, K.: Exploring simple siamese representation learning. In: Proceedings of the IEEE/CVF conference on computer vision and pattern recognition. pp. 15750--15758 (2021)

\bibitem{UFit}
Chierchia, G., Perret, B.: Ultrametric fitting by gradient descent. Advances in neural information processing systems  \textbf{32} (2019)

\bibitem{dasguptaHC}
Dasgupta, S.: A cost function for similarity-based hierarchical clustering. In: Proceedings of the Forty-Eighth Annual ACM Symposium on Theory of Computing. p. 118–127. STOC '16, Association for Computing Machinery, New York, NY, USA (2016), \url{https://doi.org/10.1145/2897518.2897527}

\bibitem{hier_index}
Deng, J., Berg, A.C., Fei-Fei, L.: Hierarchical semantic indexing for large scale image retrieval. In: CVPR 2011. pp. 785--792 (2011). \doi{10.1109/CVPR.2011.5995516}

\bibitem{imagenet}
Deng, J., Dong, W., Socher, R., Li, L.J., Li, K., Fei-Fei, L.: Imagenet: A large-scale hierarchical image database. In: 2009 IEEE Conference on Computer Vision and Pattern Recognition. pp. 248--255 (2009). \doi{10.1109/CVPR.2009.5206848}

\bibitem{graph_embd}
Epasto, A., Perozzi, B.: Is a single embedding enough? learning node representations that capture multiple social contexts. In: The World Wide Web Conference. p. 394–404. WWW '19, Association for Computing Machinery, New York, NY, USA (2019). \doi{10.1145/3308558.3313660}

\bibitem{w-mse}
Ermolov, A., Siarohin, A., Sangineto, E., Sebe, N.: Whitening for self-supervised representation learning. In: International conference on machine learning. pp. 3015--3024. PMLR (2021)

\bibitem{big-data-analy}
Fan, J., Han, F., Liu, H.: {Challenges of Big Data analysis}. National Science Review  \textbf{1}(2),  293--314 (02 2014). \doi{10.1093/nsr/nwt032}

\bibitem{byol}
Grill, J.B., Strub, F., Altch{\'e}, F., Tallec, C., Richemond, P., Buchatskaya, E., Doersch, C., Avila~Pires, B., Guo, Z., Gheshlaghi~Azar, M., et~al.: Bootstrap your own latent-a new approach to self-supervised learning. Advances in neural information processing systems  \textbf{33},  21271--21284 (2020)

\bibitem{vectordb}
Han, Y., Liu, C., Wang, P.: A comprehensive survey on vector database: Storage and retrieval technique, challenge. arXiv preprint arXiv:2310.11703  (2023)

\bibitem{provablessl}
HaoChen, J.Z., Wei, C., Gaidon, A., Ma, T.: Provable guarantees for self-supervised deep learning with spectral contrastive loss. Advances in Neural Information Processing Systems  \textbf{34},  5000--5011 (2021)

\bibitem{maskedAE}
He, K., Chen, X., Xie, S., Li, Y., Doll{\'a}r, P., Girshick, R.: Masked autoencoders are scalable vision learners. In: Proceedings of the IEEE/CVF conference on computer vision and pattern recognition. pp. 16000--16009 (2022)

\bibitem{big-data}
Jagadish, H.V., Gehrke, J., Labrinidis, A., Papakonstantinou, Y., Patel, J.M., Ramakrishnan, R., Shahabi, C.: Big data and its technical challenges. Commun. ACM  \textbf{57}(7),  86–94 (Jul 2014). \doi{10.1145/2611567}

\bibitem{cifar100}
Krizhevsky, A., Hinton, G., et~al.: Learning multiple layers of features from tiny images  (2009)

\bibitem{SSL-TKDE}
Liu, X., Zhang, F., Hou, Z., Mian, L., Wang, Z., Zhang, J., Tang, J.: Self-supervised learning: Generative or contrastive. IEEE Transactions on Knowledge and Data Engineering  \textbf{35}(1),  857--876 (2023). \doi{10.1109/TKDE.2021.3090866}

\bibitem{BBBC013}
Ljosa, V., Sokolnicki, K.L., Carpenter, A.E.: Annotated high-throughput microscopy image sets for validation. Nature methods  \textbf{9}(7),  637--637 (2012)

\bibitem{HNSW}
Malkov, Y.A., Yashunin, D.A.: Efficient and robust approximate nearest neighbor search using hierarchical navigable small world graphs. IEEE Transactions on Pattern Analysis and Machine Intelligence  \textbf{42}(4),  824--836 (2020). \doi{10.1109/TPAMI.2018.2889473}

\bibitem{marx2013big}
Marx, V.: The big challenges of big data. Nature  \textbf{498}(7453),  255--260 (2013). \doi{10.1038/498255a}

\bibitem{gHHC}
Monath, N., Zaheer, M., Silva, D., McCallum, A., Ahmed, A.: Gradient-based hierarchical clustering using continuous representations of trees in hyperbolic space. In: Proceedings of the 25th ACM SIGKDD International Conference on Knowledge Discovery \& Data Mining. p. 714–722. KDD '19, Association for Computing Machinery, New York, NY, USA (2019). \doi{10.1145/3292500.3330997}

\bibitem{cpc}
Oord, A.v.d., Li, Y., Vinyals, O.: Representation learning with contrastive predictive coding. arXiv preprint arXiv:1807.03748  (2018)

\bibitem{spectralCL}
Tan, Z., Zhang, Y., Yang, J., Yuan, Y.: Contrastive learning is spectral clustering on similarity graph. In: The Twelfth International Conference on Learning Representations (2024), \url{https://openreview.net/forum?id=hLZQTFGToA}

\bibitem{unifiedCL}
Tian, Y.: Understanding deep contrastive learning via coordinate-wise optimization. Advances in Neural Information Processing Systems  \textbf{35},  19511--19522 (2022)

\bibitem{denoisingAE}
Vincent, P., Larochelle, H., Bengio, Y., Manzagol, P.A.: Extracting and composing robust features with denoising autoencoders. In: Proceedings of the 25th International Conference on Machine Learning. p. 1096–1103. ICML '08, Association for Computing Machinery, New York, NY, USA (2008), \url{https://doi.org/10.1145/1390156.1390294}

\bibitem{JULE}
Yang, J., Parikh, D., Batra, D.: Joint unsupervised learning of deep representations and image clusters. In: Proceedings of the IEEE conference on computer vision and pattern recognition. pp. 5147--5156 (2016)

\bibitem{SeLa}
YM., A., C., R., A., V.: Self-labelling via simultaneous clustering and representation learning. In: International Conference on Learning Representations (2020), \url{https://openreview.net/forum?id=Hyx-jyBFPr}

\bibitem{barlow}
Zbontar, J., Jing, L., Misra, I., LeCun, Y., Deny, S.: Barlow twins: Self-supervised learning via redundancy reduction. In: International conference on machine learning. pp. 12310--12320. PMLR (2021)

\bibitem{vldb}
Zhang, T.: Autonomous hierarchical storage management via reinforcement learning. Proceedings of the VLDB Endowment. ISSN  \textbf{2150}, ~8097 (2024)

\bibitem{eswa}
Zhang, T., Gupta, A., Rodríguez, M.A.F., Spjuth, O., Hellander, A., Toor, S.: Data management of scientific applications in a reinforcement learning-based hierarchical storage system. Expert Systems with Applications  \textbf{237},  121443 (2023), \url{https://doi.org/10.1016/j.eswa.2023.121443}

\bibitem{hsmrl}
Zhang, T., Hellander, A., Toor, S.: Efficient hierarchical storage management empowered by reinforcement learning. IEEE Transactions on Knowledge and Data Engineering  \textbf{35}(6),  5780--5793 (2022). \doi{10.1109/TKDE.2022.3176753}

\bibitem{HAKE}
Zhang, Z., Cai, J., Zhang, Y., Wang, J.: Learning hierarchy-aware knowledge graph embeddings for link prediction. Proceedings of the AAAI Conference on Artificial Intelligence  \textbf{34}(03),  3065--3072 (Apr 2020). \doi{10.1609/aaai.v34i03.5701}

\bibitem{zhang-etal-2018-knowledge}
Zhang, Z., Zhuang, F., Qu, M., Lin, F., He, Q.: Knowledge graph embedding with hierarchical relation structure. In: Riloff, E., Chiang, D., Hockenmaier, J., Tsujii, J. (eds.) Proceedings of the 2018 Conference on Empirical Methods in Natural Language Processing. pp. 3198--3207. Association for Computational Linguistics, Brussels, Belgium (Oct-Nov 2018). \doi{10.18653/v1/D18-1358}

\end{thebibliography}
